\documentstyle[aps,preprint,epsf,floats]{revtex}

\newcommand{\nn}{\nonumber}

\def\Dsl{\hbox{/\kern-.6000em D}} 

\def\dsl{\,\raise.15ex\hbox{/}\mkern-13.5mu D}

\def\ltap{\ \raise.3ex\hbox{$<$\kern-.75em\lower1ex\hbox{$\sim$}}\ }
\def\gtap{\ \raise.3ex\hbox{$>$\kern-.75em\lower1ex\hbox{$\sim$}}\ }
\def\OMIT#1{}

\def\rmd{{\rm d}}

\begin{document}
\setlength\baselineskip{17pt}

\preprint{\tighten  \hbox{UCSD/PTH 00-14}
}

\title{The renormalization group for correlated scales: \\
one-stage versus two-stage running}

\author{Aneesh V.\ Manohar, Joan Soto, and Iain W.\ Stewart \\[4pt]}
\address{\tighten Department of Physics, University of California at San
Diego,\\[2pt] 9500 Gilman Drive, La Jolla, CA 92099  }

\maketitle

{\tighten
\begin{abstract}

Nonrelativistic bound states have two low energy scales, a soft scale $\mu_S$
of order $mv$ and an ultrasoft scale $\mu_U$ of order $mv^2$. In two-stage
running, the soft and ultrasoft scales are lowered from $m$ to $mv$, and then
the ultrasoft scale is lowered from $mv$ to $mv^2$. In one-stage running, the
two scales are lowered in a correlated way using a subtraction velocity. We
compare these two methods of summing logarithms and show that only the
correlated running in velocity space of the one-stage method correctly
reproduces the logarithms in nonrelativistic bound states in QED. The argument
for one-stage running is general, and should apply to any system with correlated
scales.

\end{abstract}
\pacs{} }

Nonrelativistic bound states have two important scales, a soft scale $\mu_S$ of
order $mv$ and an ultrasoft scale $\mu_U$ of order $mv^2$, which are of order
the typical momentum and energy of the bound state~\cite{Caswell,BBL}. In a
Coulombic bound state such as Hydrogen, the typical velocity $v$ is of order the
coupling constant $\alpha$. Logarithms of $v$ or equivalently $\alpha$ can be
summed using renormalization group equations (RGE)~\cite{LMR,amis1,amis4}.  This
is accomplished by constructing an effective theory below the scale $m$ that
contains soft and ultrasoft
modes~\cite{Caswell,BBL,LMR,amis1,amis4,Labelle,lm,Manohar,gr,Pineda,Beneke,Gries,Brambilla,KP,BenekeConf,PSH,PSee},
and then scaling the coefficients in the effective Lagrangian to low energies
using the RGE. In a properly constructed effective theory, all the large
logarithms should arise from renormalization group running, and there should be
no large logarithms in the matching conditions.  In QCD the summation of
logarithms is important because $\alpha_s \ln v \sim 1$ and the RGE
systematically accounts for the running of $\alpha_s$. In QED, $\alpha \ln
\alpha$ is small but higher order corrections are necessary to confront
precision experiments.

\begin{figure}
 \centerline{\hbox{\epsfxsize=10cm\epsfbox{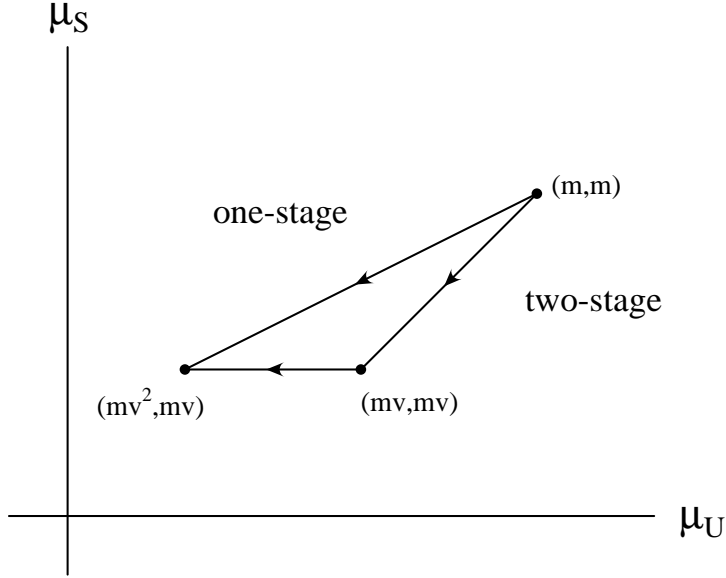}}}
 \caption{Paths in the $\mu_U-\mu_S$ plane for one-stage and
 two stage running\label{fig:path} }
\end{figure}

In two-stage running, one first lowers the subtraction scale $\mu$ from $m$ to
$mv$. At this point the soft modes, with energy-momentum of order $mv$, are
integrated out of the theory. The ultrasoft modes are then lowered from $mv$ to
$mv^2$. This is the definition of two-stage running that we will use in this
article.\footnote{\tighten See, for example, Sec.~1.4 of
Ref.~\cite{BenekeConf}. The running in Ref.~\cite{BBL} corresponds to the $m$ to
$mv$ running of the two-stage method.}  In one-stage running~\cite{LMR}, one
introduces a subtraction velocity $\nu$, with $\mu_S=m\nu$ and $\mu_U=m\nu^2$.
The subtraction velocity $\nu$ is then lowered from $\nu=1$ to $\nu=v$, so that
$\mu_S=mv$ and $\mu_U=mv^2$ at the end of the renormalization group evolution.
The difference between the two methods can be seen in Fig.~\ref{fig:path}, where
the path in the $\mu_S-\mu_U$ plane is shown. In contrasting the two methods we
will consider NRQED, where the analysis is simplified by the fact that the
coupling constant $\alpha$ does not run below the electron mass. Both the
two-stage and one-stage methods correctly reproduce observables with a single
$\ln\alpha$, such as the $\alpha^5 \ln\alpha$ Lamb shifts in Hydrogen and
positronium~\cite{PSH,PSee,amis4}. In Ref.~\cite{amis4} one-stage running was
shown to correctly reproduce the $\alpha^7 \ln^2 \alpha$ hyperfine splittings
for Hydrogen, positronium and muonium~\cite{karshenboim,kinoshita,melnikov}, the
$\alpha^3\ln^2 \alpha$ ortho and para-positronium
widths~\cite{karshenboim,melnikov,thesis}, and the $\alpha^8\ln^3\alpha$
Hydrogen Lamb shift~\cite{karshenboim}.\footnote{\tighten
Note that the $\alpha^8\ln^3\alpha$ term has also been calculated numerically in
Refs.~\cite{ms,goidenko,yerokhin} and there is general agreement on the value of
the 4-loop diagrams calculated analytically by Karshenboim in
Ref.~\cite{karshenboim}. However, Refs.~\cite{ms,yerokhin} have an additional
contribution from another graph.} This is a highly non-trivial check of
the one-stage method, and is the first time higher-order logarithms in QED bound
states were computed using the renormalization group.  In previous computations,
higher order logarithms were obtained by direct computation of multi-loop
diagrams~\cite{karshenboim,kinoshita,melnikov,thesis,kniehl}.  In this article, we explain why the two-stage
method fails to reproduce the $\ln^2\alpha$ and $\ln^3\alpha$ terms, and why the
one-stage method is essential for correctly summing all the logarithms using the
renormalization group equations.

A loop graph computed in the effective theory below $m$ typically contains
logarithms of the form $\ln p/\mu_S$, $\ln E/\mu_U$ and $\log \sqrt{m E}/\mu_S$,
where $E$ and $p$ are the typical energy and momentum in the bound state. We
will define the soft and ultrasoft anomalous dimensions $\gamma^S$ and
$\gamma^U$ by differentiating with respect to $\mu_S$ and $\mu_U$
respectively. Introducing both $\mu_S$ and $\mu_U$ allows us to simultaneously
discuss the one and two-stage running. In the effective theory the potential is
the Wilson coefficient of a four-fermion operator and has an expansion in $v$.
The coefficients $\{U_i\}$ in the potential have anomalous dimensions:
\begin{eqnarray} \label{gamSU}
 \mu_S{\rmd U_i\over \rmd \mu_S}=\gamma^{S}_i,\qquad
 \mu_U{\rmd U_i\over \rmd \mu_U}=\gamma^{U}_i,\qquad
\end{eqnarray}
where $\gamma^{S}$ and $\gamma^{U}$ are functions of $\alpha$ and potential
coefficients $\left\{U_j\right\}$.

In two-stage running, one sets $\mu=\mu_S=\mu_U$ and lowers $\mu$ from $m$ to $m
v$. Thus, one runs the potentials between $m$ and $mv$ using the anomalous
dimension $\Gamma^S$,\footnote{\tighten In pNRQED, $\Gamma^S_i$ is called the
soft anomalous dimension because it governs the running between the hard scale
$m$ and the soft scale $mv$. In the one-stage approach, it is $\gamma^S_i$ which
is called the soft anomalous dimension. This is merely a difference in
terminology with no physical consequences.}
\begin{eqnarray} \label{g1}
  \mu{\rmd U_i\over \rmd \mu}=\Gamma^S_i=\gamma^{S}_i+\gamma^{U}_i \qquad\qquad
  mv < \mu < m \,.
\end{eqnarray}
One then integrates out the soft modes at $\mu=mv$, and runs to $mv^2$ with the
ultrasoft anomalous dimension
\begin{eqnarray} \label{g2}
  \mu{\rmd U_i\over \rmd \mu}=\gamma^{U}_i\phantom{+\gamma^{U}_i} \qquad\qquad
  mv^2 < \mu < mv \,.
\end{eqnarray}
In one-stage running the subtraction scales $\mu_S$ and $\mu_U$ are
different,\footnote{\tighten In dimensional regularization where $d=4-2\epsilon$,
factors of $\mu_S^\epsilon$ and $\mu_U^\epsilon$ appear along with interaction
vertices involving soft and ultrasoft modes respectively \cite{LMR,amis1}.}
$\mu_S=m\nu$ and $\mu_U=m\nu^2$.  With the velocity renormalization group one
runs from $\nu=1$ to $\nu=v$ with the anomalous dimension
\begin{eqnarray} \label{gv}
 \nu{\rmd U_i\over \rmd \nu}=\gamma^{S}_i+2 \gamma^{U}_i \qquad\qquad
  v < \nu < 1 \,.
\end{eqnarray}

As an example, consider the running of the $\delta$-function contact potential
for Hydrogen in the limit $m_{\rm proton}\to\infty$. The coefficient of this
potential will be defined as $U_2$, using the notation of
Ref.~\cite{amis4}. $U_2$ plays a key role in NRQED since it is the lowest order
coefficient that runs. With one-stage running $U_2$ has a velocity
renormalization group equation of the form~\cite{amis4}
\begin{eqnarray} \label{U2vrge}
  \nu {\rmd U_2 \over \rmd \nu}=2 \alpha^2 \lambda^U + \alpha \lambda^S U_2^2
  + \ldots \,.
\end{eqnarray}
Here $\lambda^U = -{4 /(3 m^2)}$, $\lambda^S = {m^2/\pi}$, and the matching
value $U_2(\nu=1)=\pi\alpha/(2m^2)$, where $m$ is the electron mass. The
coefficient $U_2$ is of order $\alpha$, so the first term in Eq.~(\ref{U2vrge})
is the leading order (LO) anomalous dimension, and the second term belongs to
the next-to-leading order (NLO) anomalous dimension.  For Hydrogen, the first
term is purely ultrasoft and the second term is purely soft, so
\begin{eqnarray} \label{gammaus}
  \gamma^U = \alpha^2 \lambda^U,\qquad \qquad
  \gamma^S = \alpha \lambda^S U_2^2 \,.
\end{eqnarray}
In Eq.~(\ref{U2vrge}), we have only retained the NLO term relevant for the
$\alpha^8 \ln^3 \alpha$ Lamb shift in Hydrogen. The full equation is given in
Ref.~\cite{amis4} and gives additional contributions to $\alpha^7 \ln^2\alpha$
and $\alpha^6 \ln\alpha$, but the simplified version in Eq.~(\ref{U2vrge}) will
suffice for this paper.

In our example, integrating the renormalization group equation using the LO
anomalous dimension and one-stage running gives
\begin{eqnarray} \label{U2lo}
  U_2(\nu=v) = U_2(1) + 2 \alpha^2 \lambda^U \ln v \,,
\end{eqnarray}
which is the leading order series $\alpha^{n+1} \ln^n v$ for $U_2(v)$ and
determines the $\alpha^5\ln\alpha$ Lamb shift correction.  There is only a
single-log term since the LO anomalous dimension is a constant.  To
next-to-leading log order only the leading log value of $U_2$ is needed on the
RHS of Eq.~(\ref{U2vrge}). Substituting Eq.~(\ref{U2lo}) for $U_2$ into
Eq.~(\ref{U2vrge}) and reintegrating gives
\begin{eqnarray} \label{rlt1}
 U_2(\nu=v) &=& U_2(1) + 2 \alpha^2 \lambda^U \ln v \nn \\ && + \alpha \lambda^S
 U_2(1)^2 \ln v + 2\,\alpha^3 \lambda^U \lambda^S U_2(1) \ln^2 v + {4\over 3}
 \alpha^5 \left(\lambda^U\right) ^2 \lambda^S \ln^3 v \,,
\end{eqnarray}
which includes the LO series $\alpha^{n+1} \ln^n v$ and the next-to-leading
order series $\alpha^{n+2} \ln^n v$. The highest logarithm in the NLO series is
$\ln^3 v$, since the NLO anomalous dimension has a $U_2^2$ term that contains
$\ln^2 v$ terms when the LO running for $U_2$ is used.  The $\ln^3 v$ term gives
an $\alpha^8\ln^3\alpha$ Lamb shift for Hydrogen in agreement with
Karshenboim~\cite{karshenboim}.

Next we redo this example using two-stage running. In the first step, one
integrates
\begin{eqnarray}\label{rge1}
 \mu {\rmd U_2\over \rmd \mu}=\alpha^2 \lambda^U +  \alpha \lambda^S U_2^2 \,,
\end{eqnarray}
to give
\begin{eqnarray} \label{ans1}
 U_2(\mu=mv) &=& U_2(1) + \alpha^2 \lambda^U \ln v  \nn\\
 &&+ \alpha \lambda^S U_2(1)^2
  \ln v +  \alpha^3 \lambda^U \lambda^S  U_2(1)  \ln^2 v +
 {1\over 3} \alpha^5 \left(\lambda^U\right) ^2  \lambda^S \ln^3 v \,,
\end{eqnarray}
where $U_2(1)=U_2(\mu=m)$. In the second step one integrates
\begin{eqnarray}\label{rge2}
 \mu {\rmd U_2\over \rmd \mu}=\alpha^2 \lambda^U.
\end{eqnarray}
Since the $\lambda^S$ term comes solely from the soft scale it does not appear
in the RGE for the second step.  Integrating Eq.~(\ref{rge2}) using
Eq.~(\ref{ans1}) as the starting point gives the final result
\begin{eqnarray} \label{rlt2}
 U_2(\mu=mv^2) &=& U_2(1) + 2 \alpha^2 \lambda^U \ln v \nn \\ && +
 \alpha\lambda^S U_2(1)^2 \ln v + \alpha^3 \lambda^U \lambda^S U_2(1)
 \ln^2 v + {1\over 3} \alpha^5 \left(\lambda^U\right) ^2 \lambda^S \ln^3 v \,.
\end{eqnarray}
Comparing Eqs.~(\ref{rlt1}), and Eqs.~(\ref{rlt2}), we see that the final
expressions agree in the single-log term, but disagree in the higher-order
logarithms.

One can study the difference between one- and two-stage running in the general
case, by integrating the renormalization group equations perturbatively. The
single-log terms (i.e. proportional to $\ln v$) are obtained by integrating the
renormalization group equations using the values for the anomalous dimensions at
the matching scale, i.e.\ by substituting the tree-level values for $\left\{
U_i\right\}$ in $\gamma\left(\left\{U_i\right\} \right)$, which will be denoted
by $\gamma(0)$.  In the two-stage approach, one finds that
\begin{eqnarray}  \label{one}
 U_i(\mu=mv^2)= \Gamma^{S}_i(0) \ln v +
  \gamma^{U}_i(0)  \ln v=\left[ \gamma^{S}_i(0)+\gamma^{U}_i(0) \right] \ln v +
  \gamma^{U}_i(0)  \ln v ,
\end{eqnarray}
where the first term is the contribution between $m$ and $mv$, and the second
term is the contribution between $mv$ and $mv^2$.  In the one-stage approach,
one finds
\begin{eqnarray}\label{two}
  U_i(\nu=v)=\left[ \gamma^{S}_i(0)+2\gamma^{U}_i(0) \right] \ln v .
\end{eqnarray}
Both methods give the same answer for the $\ln v$ term, since
\begin{eqnarray}\label{true}
 \Gamma^{S}_i(0) + \gamma^{U}_i(0) =  \left[ \gamma^{S}_i(0)+\gamma^{U}_i(0)
 \right]+\gamma^U_i(0)
 = \left[ \gamma^{S}_i(0)+2\gamma^{U}_i(0) \right].
\end{eqnarray}
This is to be expected, since the anomalous dimensions in both methods should
reproduce the single logarithms in the graphs from which they were calculated.
This is true of our Hydrogen example, cf. Eq.~(\ref{rlt1}) and Eq.~(\ref{rlt2}).
We have also checked that this is true for the $\alpha^5 \ln\alpha$ Lamb shift
for positronium in which case the LO anomalous dimension has both an ultrasoft
and soft contribution.

The non-trivial terms are the higher order logarithms, which are generated when
one includes the running potentials in the expression for the anomalous
dimensions $\gamma^{S,U}$. The anomalous dimensions depend on $t=\ln v$ through
their dependence on $U_i$. Consider expanding $\gamma$ in powers of $t$,
\begin{eqnarray}
 \gamma_j\left(t\right)&=&\gamma_j\left(0\right) + t
 \left( {{\rm d} \gamma_j\over {\rm d} t}\right)_{t=0}  + {t^2 \over 2}
 \left( {{\rm d}^2 \gamma_j\over {\rm d} t^2}\right)_{t=0}  + \ldots,
\end{eqnarray}
where the derivatives with respect to $t$ can be evaluated using the chain rule,
\begin{eqnarray} \label{expnG}
 \gamma_j\left(t\right)&=&\gamma_j\left(0\right) + t
 \left( {\partial \gamma_j\over \partial U_k} {{\rm d} U_k \over {\rm d} t}
 \right)_{t=0}  + {t^2 \over 2} \left( {\partial^2 \gamma_j\over \partial U_k
 \partial U_l} {{\rm d} U_k \over {\rm d} t} {{\rm d} U_l \over {\rm d} t} +
 {\partial \gamma_j \over \partial U_k}{{\rm d}^2 U_k \over {\rm d} t^2}
 \right)_{t=0} + \ldots \nn \\[5pt]
 &=&\gamma_j\left(0\right) + t
 \left( {\partial \gamma_j\over \partial U_k} {\gamma_k}
 \right)_{t=0}  + {t^2 \over 2} \left( {\partial^2 \gamma_j\over \partial U_k
 \partial U_l} {\gamma_k} {\gamma_l} +
 {\partial \gamma_j \over \partial U_k}{\partial \gamma_k \over \partial U_l}
 {\gamma_l}  \right)_{t=0} + \ldots.
\end{eqnarray}
For Hydrogen, the $\ln^2 v$ and $\ln^3 v$ terms are generated from cross-terms
of the form $\gamma^U\, \partial\gamma^S/\partial U_2$ and
$\left(\gamma^U\right)^2\, \partial^2 \gamma^S/\partial U_2^2$ that involve both
soft and ultrasoft anomalous dimensions. In the one-stage approach, the
renormalization group equations are integrated with the anomalous dimension in
Eq.~(\ref{gv}) between $\nu=1$ and $\nu=v$, so the cross-terms have the
structure
\begin{eqnarray} \label{1st}
 (2\gamma^U)\: {\partial\gamma^S \over \partial U_2}\: \ln^2 v \,,\qquad\qquad
 (2\gamma^U)^2\: {\partial^2 \gamma^S \over \partial U_2^2}\:  \ln^3 v  .
\end{eqnarray}
In the two-stage approach, one integrates Eq.~(\ref{g1}) from $m$ to $mv$, after
which $\gamma^S=0$, so the corresponding terms have the structure
\begin{eqnarray} \label{2st}
 (\gamma^U)\: {\partial\gamma^S \over \partial U_2}\: \ln^2 v ,\qquad\qquad
 (\gamma^U)^2\: {\partial^2 \gamma^S \over \partial U_2^2}\: \ln^3 v ,
\end{eqnarray}
which differ by a factor of two and four, respectively, from the one-stage
running values.\footnote{\tighten Recall that there are terms in the NLO
anomalous dimension that are not shown in Eq.~(\ref{U2vrge}) which contribute to
terms involving $\ln^2\alpha$ in the energy. Including these additional terms
makes the first term in Eq.~(\ref{1st}) and in Eq.~(\ref{2st}) more complicated,
however the one- and two-stage $\ln^2 v$ results still differ by an overall
factor of $2$.}

In general, one-stage and two-stage running give different values for the final
answer. Define the anomalous dimension vector
$\mbox{\boldmath$\gamma$}=(\gamma^U,\gamma^S)$. The running is given by
integrating the anomalous dimension vector along the path shown in
Fig.~\ref{fig:path}. The integral is path dependent if $\mbox{\boldmath$\nabla$}
\times \mbox{\boldmath$\gamma$} \not= \mathbf{0}$, i.e.
\begin{eqnarray}
 \mu_U {\rmd \gamma^S \over \rmd \mu_U} \not = \mu_S {d \gamma^U \over d \mu_S}.
\end{eqnarray}
In our Hydrogen example, $\gamma^U=\alpha^2 \lambda^U$ and $\gamma^S =
\alpha \lambda^S U_2^2$, so that
\begin{eqnarray}
\mu_U {\rmd \gamma^S \over \rmd \mu_U} &=&  2 \alpha\, \lambda^S\, U_2\:
\mu_U {\rmd U_2 \over \rmd \mu_U} = 2 \alpha^3 \lambda^S \lambda^U U_2, \nn \\
\mu_S {\rmd \gamma^U \over \rmd \mu_S} &=& 0,
\end{eqnarray}
and the integral is path-dependent.

\begin{figure}
 \centerline{\hbox{\epsfxsize=6cm\epsfbox{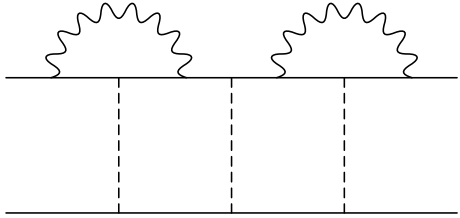}}} \caption{A four-loop
 graph that contributes to the $\alpha^8 \ln^3 \alpha$ Lamb shift in
 Hydrogen.\label{fig:four} }
\end{figure}

In Ref.~\cite{amis4}, one-stage running was shown to correctly give the
$\alpha^8 \ln^3 \alpha$ Lamb shift and $\alpha^7 \ln^2 \alpha$ hyperfine
splitting for Hydrogen, muonium and positronium, as well as the $\alpha^3 \ln^2
\alpha$ correction to the decay width for positronium. Since the answer differs
from two-stage running, this implies that two-stage running does not sum all the
logarithms of $v$. The reason can be seen by considering a diagram with mixed
soft and ultrasoft divergences, such as the four-loop graph in
Fig.~\ref{fig:four}.  If we calculate logarithms of $\alpha$ directly by
calculating matrix elements (using vertices at the hard scale), then this graph
contributes to the $\alpha^8 \ln^3 \alpha$ Lamb shift.
Calculating either ultrasoft subgraph in Fig.~\ref{fig:four} gives a term of the
form $\ln (k^0/\mu_U)$, whereas the two-loop potential subgraph gives a term of
the form $\ln (\sqrt{m k^0}/\mu_S)$, where $k^0$ is the time-component of a loop
momentum. In the RGE, to make sure that both logarithms are small it is
essential that $\mu_U = \mu_S^2/m$, so that the soft and ultrasoft scales are
correlated. If one attempts to set $\mu_S=\mu_U=\mu$, then one cannot
simultaneously minimize both soft and ultrasoft logarithms inside this four-loop
graph. In the two-stage trajectory shown in Fig.~\ref{fig:path}, the logarithms
$\ln (k^0/\mu_U)$ and $\ln (\sqrt{m k^0}/\mu_S)$ are made small at the endpoint
of the path. However, both logarithms are not small everywhere along the
integration path, so that the logarithms are not minimized inside subgraphs of
the full graph. As a result, two-stage running gives the correct single log, but
not the higher order logarithms. In the one-stage approach the logarithms are
small along the entire integration path and the higher order logarithms are
correctly reproduced.

Although two-stage running seems natural within the pNRQED
framework~\cite{Pineda,PSH,PSee}, a single stage running can be implemented as
well. The discussion above suggests that even though matching from QED to pNRQED
can be done in a two-stage procedure through NRQED, the renormalization group
improvement should be done in one stage. However, it may well be that a more
sophisticated version of two-stage running does properly take into account the
fact that the soft and ultrasoft scales are correlated.

The difference between one- and two-stage running is a generic feature of
systems with correlated scales. In nonrelativistic bound states, the soft and
ultrasoft scales are determined in terms of the velocity $v$, with $\mu_S=mv$
and $\mu_U=mv^2$, and one needs to use one-stage running in velocity space.  We
have illustrated this point using QED, but the arguments are obviously also
valid for weak coupling nonrelativistic QCD bound states.  Correlated scales
also occur in the problem of Sudakov logarithms near kinematic endpoints (for
effective theory approaches see Refs.~\cite{sudakov}). In $B\to X_s \gamma$
decay, there are two correlated scales, the collinear scale $m_b\sqrt{1-x}$ and
the soft scale $m_b(1-x)$ (here $x=2E_\gamma/m_b$) that are important as $x\to
1$. To sum subleading Sudakov logarithms using the renormalization group one
needs to run in $x$ (or equivalently moment) space. One-stage running might also
be applicable to finite temperature field theory, where one has the correlated
scales $T$, $gT$ and $g^2T$. This will be discussed elsewhere.

This work was supported in part by the Department of Energy under grant
DOE-FG03-97ER40546. I.S. was also supported in part by the National Science
Foundation under award NYI PHY-9457911 and by NSERC of Canada. J.S. is also
supported by the AEN98-031 (Spain), the 1998SGR 00026 (Catalonia) and the BGP-08
fellowship (Catalonia). He thanks A. Manohar and the high energy physics group
at UCSD for their warm hospitality.

{\tighten

} 

\end{document}